\documentclass[final,5p,times,twocolumn,nopreprintline]{elsarticle}
\usepackage{amsmath,amssymb,amsfonts}
\usepackage{graphicx}
\usepackage{slashed}
\usepackage{booktabs,tabulary}

\newcommand{\beq}{\begin{equation}}
\newcommand{\eeq}{\end{equation}}
\newcommand{\beqa}{\begin{eqnarray}}
\newcommand{\eeqa}{\end{eqnarray}}

\newcommand{\mpi}{M_{\pi}}
\newcommand{\mpii}{M_{\pi^0}}

\newcommand{\diff}{\text{d}}

\newcommand{\F}{\mathcal{F}}

\renewcommand{\Im}{\text{Im}\,}

\newcommand{\GeV}{\,\text{GeV}}

\renewcommand{\arraystretch}{1.2}

\allowdisplaybreaks[1]

\begin{document}

\renewcommand{\theequation}{\arabic{equation}}

\begin{frontmatter}

\title{Towards a data-driven analysis of hadronic light-by-light scattering}

\author[Bern]{Gilberto Colangelo}
\author[Bern]{Martin Hoferichter}
\author[Bonn]{Bastian Kubis}
\author[Bern]{Massimiliano Procura}
\author[Bern]{Peter Stoffer}

\address[Bern]{Albert Einstein Center for Fundamental Physics,
Institute for Theoretical Physics,
Universit\"at Bern, Sidlerstrasse 5, CH--3012 Bern, Switzerland}
\address[Bonn]{Helmholtz-Institut f\"ur Strahlen- und Kernphysik (Theorie) 
   and Bethe Center for Theoretical Physics, Universit\"at Bonn, D--53115 Bonn, Germany}

\begin{abstract}
  The hadronic light-by-light contribution to the anomalous magnetic moment
  of the muon was recently analyzed in the framework of dispersion theory,
  providing a systematic formalism where all input quantities are expressed
  in terms of on-shell form factors and scattering amplitudes that are in
  principle accessible in experiment. We briefly review the main ideas
  behind this framework and discuss the various experimental ingredients
  needed for the evaluation of one- and two-pion intermediate states. In
  particular, we identify processes that in the absence of data for
  doubly-virtual pion--photon interactions can help constrain parameters in
  the dispersive reconstruction of the relevant input quantities, the pion
  transition form factor and the helicity partial waves for
  $\gamma^*\gamma^*\to\pi\pi$.
\end{abstract}


\end{frontmatter}

\section{Introduction}

The limiting factor in the accuracy of the Standard-Model prediction for
the anomalous magnetic moment of the muon $a_\mu=(g-2)_\mu/2$ is control
over hadronic uncertainties~\cite{JN,Prades:2009tw}.  The leading hadronic
contribution, hadronic vacuum polarization, is related to the total
hadronic cross section in $e^+e^-$ annihilation, so that the improvements
necessary to compete with the projected accuracy of the FNAL and J-PARC
experiments can be achieved with a dedicated $e^+e^-$ program, see
e.g.~\cite{g-2wp,Benayoun:2014tra}.  Owing to the complexity of the hadronic light-by-light
(HLbL) tensor, a similar data-driven approach for the
subleading\footnote{At this order also two-loop diagrams with insertions of
  hadronic vacuum polarization appear~\cite{Calmet:1976kd}. Even
  higher-order hadronic contributions have been recently considered
  in~\cite{Kurz:2014wya,Colangelo:2014qya}.}  HLbL scattering contribution
has only recently been suggested, and only for the leading hadronic
channels~\cite{CHPS}. In contrast to previous approaches~\cite{deRafael:1993za,Bijnens:1995cc,BPP95,Bijnens:2001cq,Hayakawa:1995ps,Hayakawa:1996ki,Hayakawa:1997rq,Knecht:2001qg,KN,RamseyMusolf:2002cy,MV,Goecke:2010if,Roig:2014uja}, this formalism aims at providing a direct link between data and the HLbL contribution to $a_\mu$.
An alternative strategy to reduce model-dependence in HLbL relies on lattice QCD, see~\cite{Blum:2014oka} for a first calculation.

The dispersive framework in~\cite{CHPS} includes both the dominant
pseudoscalar-pole contributions as well as two-meson intermediate states,
thus covering the most important channels.
In view of the fact that a data-driven approach
for the HLbL contribution is substantially more involved than that for
HVP, we present here an overview of this approach leaving aside all
theoretical details, and emphasize which measurements can help constrain
the required hadronic input. At present such an overview can only be
obtained after studying several different theoretical papers.
It is, however, essential that also experimentalists become fully aware that
some measurements may have a substantial and model-independent impact on
a better determination of the HLbL contribution to $a_\mu$. This is the
main aim of the present letter.

\section{Theoretical framework}

\subsection{Dispersion relations}

In dispersion theory the matrix element of interest is reconstructed from information on its analytic structure: residues of poles, discontinuities along cuts, and subtraction constants (representing singularities at infinity). In contrast to HVP, the complexity of the HLbL tensor prohibits the summation of all possible intermediate states into a single dispersion relation. Instead, one has to rely on an expansion in the mass of allowed intermediate states, justified by higher thresholds and phase-space suppression in the dispersive integrals. In this paper we concentrate on the lowest-lying intermediate states, the $\pi^0$ pole and $\pi\pi$ cuts, that illustrate the basic features of our dispersive approach and are expected to be most relevant numerically. We will comment on higher intermediate states in Sect.~\ref{sec:higher_IS}.

\begin{figure}
\begin{center}
\raisebox{0.65cm}{\includegraphics[width=0.306\linewidth,clip]{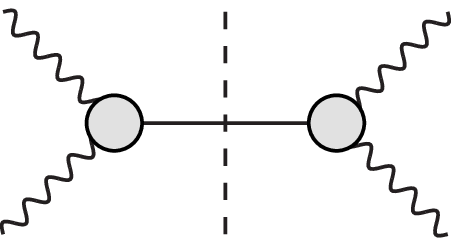}}\quad
\includegraphics[width=0.306\linewidth,clip]{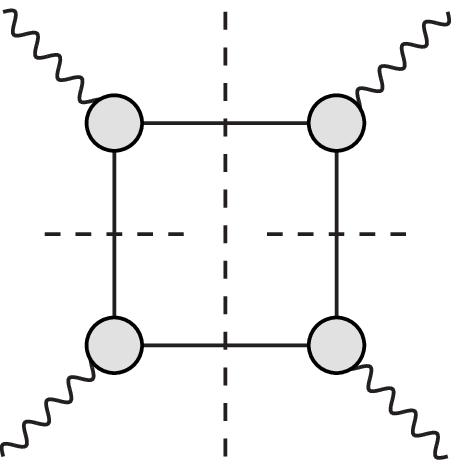}\quad
\raisebox{0.65cm}{\includegraphics[width=0.306\linewidth,clip]{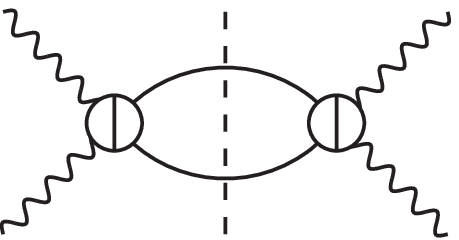}}
\end{center}
\caption{Representative unitarity diagrams for the pion pole (left), the FsQED contribution (middle), and $\pi\pi$ rescattering (right). The gray blobs refer to the pertinent pion form factors, those with vertical line to the non-pole $\gamma^*\gamma^*\to\pi\pi$ amplitude. The dashed lines indicate the cutting of pion propagators. For more details see~\cite{CHPS}.}
\label{fig:unitarity_diagrams}
\end{figure}

Given that each contribution to the HLbL tensor is uniquely defined by its analytic structure, it can be related unambiguously to 
a certain physical intermediate state. We decompose the HLbL tensor according to
\beq
\label{eq:Pibreakdown}
\Pi_{\mu\nu\lambda\sigma}=\Pi_{\mu\nu\lambda\sigma}^{\pi^0}
+\Pi_{\mu\nu\lambda\sigma}^{\text{FsQED}}+\Pi_{\mu\nu\lambda\sigma}^{\pi\pi}+\cdots,
\eeq
where $\Pi_{\mu\nu\lambda\sigma}^{\pi^0}$ denotes the pion
pole, $\Pi_{\mu\nu\lambda\sigma}^{\text{FsQED}}$ the amplitude in scalar QED
with vertices dressed by the pion vector form
factor $F_\pi^V$ (FsQED), $\Pi^{\pi\pi}_{\mu\nu\lambda\sigma}$ includes the remaining
$\pi\pi$ contribution, and the ellipsis higher intermediate states. Representative unitarity diagrams for each 
term are shown in Fig.~\ref{fig:unitarity_diagrams}.

The separation of the FsQED amplitude ensures that contributions with simultaneous cuts in two 
kinematic variables are correctly accounted for. In fact,
$\Pi_{\mu\nu\lambda\sigma}^{\text{FsQED}}$ is completely fixed by the pion vector form factor, 
see~\cite{CHPS} for details and explicit expressions. Since for this purpose $F_\pi^V$ is known to sufficient accuracy experimentally,
$\Pi_{\mu\nu\lambda\sigma}^{\text{FsQED}}$ is completely determined and we will 
concentrate on reviewing the central results for $\Pi_{\mu\nu\lambda\sigma}^{\pi^0}$ and $\Pi_{\mu\nu\lambda\sigma}^{\pi\pi}$ in the following.

\subsection{Pion pole}
\label{sec:pion_pole}

The residue of the pion pole is determined by the pion transition form factor $\F_{\pi^0\gamma^*\gamma^*}(q_1^2,q_2^2)$. The corresponding contribution to $a_\mu$ follows from~\cite{KN} 
\begin{align}
\label{amu_pole}
 a_\mu^{\pi^0}&=-e^6\int\frac{\diff^4q_1}{(2\pi)^4}\int\frac{\diff^4q_2}{(2\pi)^4}
 \frac{1}{q_1^2q_2^2sZ_1 Z_2}\notag\\
 &\times\Bigg\{\frac{\F_{\pi^0\gamma^*\gamma^*}\big(q_1^2,q_2^2\big)\F_{\pi^0\gamma^*\gamma^*}\big(s,0\big)}{s-\mpii^2}T^{\pi^0}_1(q_1,q_2;p)
 \notag\\
 &\qquad+\frac{\F_{\pi^0\gamma^*\gamma^*}\big(s,q_2^2\big)\F_{\pi^0\gamma^*\gamma^*}\big(q_1^2,0\big)}{q_1^2-\mpii^2}T^{\pi^0}_2(q_1,q_2;p)\Bigg\},\notag\\
 Z_1&=(p+q_1)^2-m^2,\qquad Z_2=(p-q_2)^2-m^2,\notag\\
 s&=(q_1+q_2)^2,
\end{align}
where $m$ denotes the mass of the muon, $p$ its momentum, $e=\sqrt{4\pi\alpha}$ the electric charge, and the $T^{\pi^0}_i(q_1,q_2;p)$ are known kinematic functions.

It should be mentioned that the relation~\eqref{amu_pole} only represents the $\pi^0$ pole, it does not, on its own, satisfy QCD short-distance constraints. As pointed out in~\cite{MV}, the pion pole as defined in~\eqref{amu_pole} tends faster to zero for large $q^2$ than required by perturbative QCD due to the momentum dependence in the singly-virtual form factors. The correct high-energy behavior is only restored by the exchange of heavier pseudoscalar resonances, but the pion-pole contribution, by its strict dispersive definition, is unambiguously given as stated in~\eqref{amu_pole}.

\subsection{$\pi\pi$ intermediate states}

The contribution from $\pi\pi$ intermediate states can be expressed as~\cite{CHPS}
\beq
\label{amu_pipi}
 a_\mu^{\pi\pi}=e^6\int\frac{\diff^4q_1}{(2\pi)^4}\int\frac{\diff^4q_2}{(2\pi)^4}\frac{\sum_i I_i\big(s,q_1^2,q_2^2\big)T^{\pi\pi}_i\big(q_1,q_2;p\big)}{q_1^2q_2^2s Z_1 Z_2},
\eeq
in a way similar to the pion pole~\eqref{amu_pole}. The $T^{\pi\pi}_i(q_1,q_2;p)$ again denote known kinematic functions, while the information 
on the amplitude on the cut is hidden in the dispersive integrals $I_i(s,q_1^2,q_2^2)$. For instance, the first $S$-wave term reads
\begin{align}
\label{Ii_imag}
I_1\big(s,q_1^2,q_2^2\big)&=\frac{1}{\pi}\int\limits_{4\mpi^2}^\infty\frac{\diff s'}{s'-s}\bigg[\bigg(\frac{1}{s'-s}
 -\frac{s'-q_1^2-q_2^2}{\lambda\big(s',q_1^2,q_2^2\big)}\bigg)\notag\\
 &\qquad\times\Im h_{++,++}^0\big(s';q_1^2,q_2^2;s,0\big)\notag\\
 &+\frac{2\xi_1\xi_2}{\lambda\big(s',q_1^2,q_2^2\big)} \Im h^0_{00,++}\big(s';q_1^2,q_2^2;s,0\big)\bigg]
 \end{align}
 with K\"all\'en function $\lambda(x,y,z)=x^2+y^2+z^2-2(xy+xz+yz)$, normalization of longitudinal polarization vectors $\xi_i$, and partial-wave helicity amplitudes 
 $h_{\lambda_1\lambda_2,\lambda_3\lambda_4}^J(s;q_1^2,q_2^2;q_3^2,q_4^2)$ for 
 \beq
 \gamma^*\big(q_1,\lambda_1\big)  \gamma^*\big(q_2,\lambda_2\big)\to \gamma^*\big(q_3,\lambda_3\big)  \gamma^*\big(q_4,\lambda_4\big)
 \eeq
 with angular momentum $J$. By means of partial-wave unitarity 
 \begin{align}
&\Im h_{\lambda_1\lambda_2,\lambda_3\lambda_4}^J\big(s;q_1^2,q_2^2;q_3^2,q_4^2\big)\notag\\
&=\frac{\sqrt{1-4\mpi^2/s}}{16\pi}h_{J,\lambda_1\lambda_2}\big(s;q_1^2,q_2^2\big)h_{J,\lambda_3\lambda_4}\big(s;q_3^2,q_4^2\big),
\end{align}
 the imaginary part in~\eqref{Ii_imag} is related to the helicity partial waves $h_{J,\lambda_1\lambda_2}(s;q_1^2,q_2^2)$
 for $\gamma^*\gamma^*\to\pi\pi$, which have to be determined from experiment.
 
One key feature in the derivation of~\eqref{amu_pipi} concerns the subtraction polynomial. Frequently, 
dispersion relations need to be subtracted to render the integrals convergent, and the ensuing subtraction constants are free parameters of the approach that need to be determined either from experiment or by further theoretical means, such as effective field theories or lattice QCD. 
For HLbL scattering, however, gauge invariance puts very stringent constraints on the amplitude and the subtraction polynomial. Therefore, the situation is actually similar to HVP, where the combination of analyticity, unitarity, and gauge invariance provides a parameter-free relation between the contribution to $a_\mu$ and the experimental input, the hadronic $e^+e^-$ cross section, as well.

\section{Experimental input}

\begin{figure}
\begin{center}
\includegraphics[width=0.4\linewidth,clip]{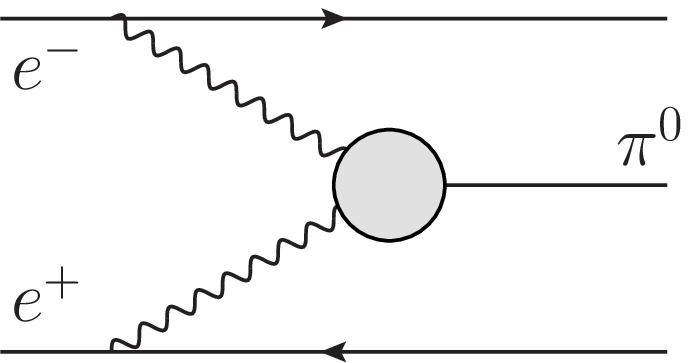}\qquad
\includegraphics[width=0.4\linewidth,clip]{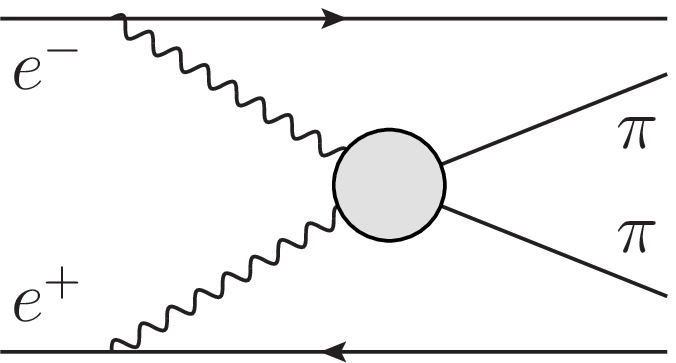}
\end{center}
\caption{$e^+e^-\to e^+e^-\pi^0$ and $e^+e^-\to e^+e^-\pi\pi$ in space-like kinematics.}
\label{fig:spacelike}
\end{figure}

\begin{figure}
\begin{center}
\includegraphics[width=0.4\linewidth,clip]{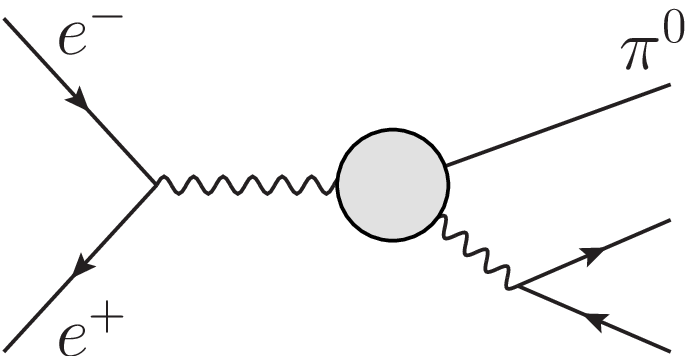}\qquad
\includegraphics[width=0.4\linewidth,clip]{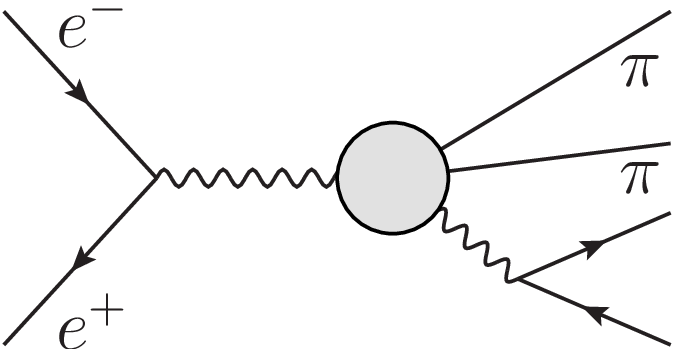}
\end{center}
\caption{$e^+e^-\to e^+e^-\pi^0$ and $e^+e^-\to e^+e^-\pi\pi$ in time-like kinematics.}
\label{fig:timelike}
\end{figure}

By means of a Wick rotation the loop integrals in~\eqref{amu_pole} and~\eqref{amu_pipi} can be brought into such a form that only space-like momenta appear in the integral, so that in principle all required information can be extracted from the processes depicted in Fig.~\ref{fig:spacelike}.
However, this would require double-tag measurements for arbitrary negative virtualities, and, in the $\pi\pi$ case, sufficient angular information to perform a partial-wave analysis.

Although such detailed information about doubly-virtual pion--photon interactions is currently not available, there are existing and planned measurements involving real or singly-virtual processes, not only in space-like but also in time-like kinematics, see Fig.~\ref{fig:timelike} for the doubly-virtual time-like case. All this information can be used to reconstruct, in turn, both the pion transition form factor as well as $\gamma^*\gamma^*\to\pi\pi$ partial waves using dispersion relations. The benefits from such a program are manifold:
first, it makes sure that the resulting input for~\eqref{amu_pole} and~\eqref{amu_pipi} is consistent with analyticity and unitarity. Second, it would allow for a global analysis of all information of pion--photon interactions from all kinematic regions. Third, it should allow for the identification of processes and kinematic regions that are responsible for the largest uncertainty in the final HLbL prediction and should therefore be subject to further experimental scrutiny. In this paper we do not yet make quantitative statements, but rather identify processes potentially relevant, as well as overlap in the calculation of the one- and two-pion input.

For the pion transition form factor some work along these lines has already been presented in~\cite{Czerwinski,NKS,HKS,SKN,MesonNet}.
Similarly, analyses of the on-shell process $\gamma\gamma\to\pi\pi$~\cite{GMM,HPS}, the singly-virtual reaction $\gamma^*\gamma\to\pi\pi$~\cite{Moussallam13}, and, some first steps, for the doubly-virtual case $\gamma^*\gamma^*\to\pi\pi$~\cite{HCPS} have been performed. In particular, in~\cite{HCPS} it was shown how to properly account for so-called anomalous thresholds~\cite{Mandelstam,LMS}, which emerge in time-like kinematics for $\gamma^*\gamma^*\to\pi\pi$ as a new feature concerning the analytic properties of the scattering amplitude.

\begin{figure*}
\begin{center}
\includegraphics[height=0.9\linewidth,angle=-90]{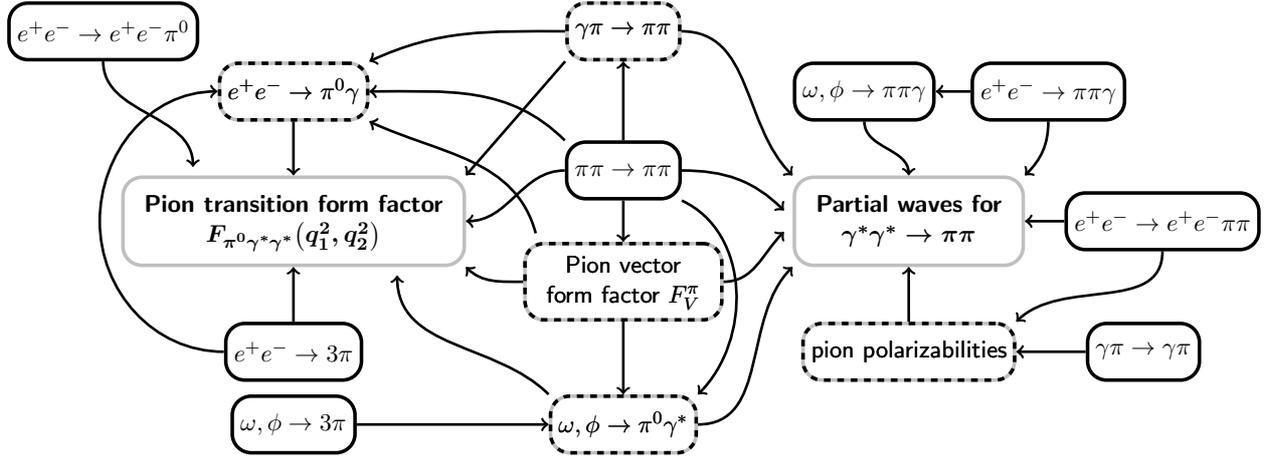}
\end{center}
\caption{Processes relevant for the dispersive reconstruction of the pion transition form factor and the helicity partial waves for $\gamma^*\gamma^*\to\pi\pi$. Gray boxes refer to the final ingredients for
$a_\mu$, black ones to quantities considered as input, and dashed boxes to
quantities that can both be measured and calculated theoretically.}
\label{fig:flowchart}
\end{figure*}

A collection of processes relevant for the execution of this program for
one- and two-pion intermediate states is shown in Fig.~\ref{fig:flowchart}.
The line coding is such that gray boxes refer to the final ingredients for
$a_\mu$, black ones to quantities considered as input, and dashed boxes to
quantities that can both be measured and calculated theoretically. The last
class of processes serves as a check of agreement between experiment and
theory at various stages: the theoretical representations are often confined
to \emph{elastic} unitarity and include at most $\pi\pi$ intermediate states,
while some quantities, such as the pion vector form factor $F_V^\pi$, are known 
experimentally to much higher precision, and at higher energies than 
accessible to the elastic approximation. In this way, the difference between the
full experimental result and the dispersive reconstruction can be taken as
indicative of the impact of higher intermediate states.

The crucial role of elastic unitarity is also a manifestation of the fact that
by definition the dispersive formalism works best at low energies, where only a limited number
of intermediate states contribute. Due to the energy denominators (and phase-space suppression)
this is precisely the energy region most relevant in the HLbL integrals,
see~\eqref{amu_pole}, \eqref{amu_pipi}, and \eqref{Ii_imag}. 
Therefore, while high-energy data will be highly welcome when it comes to addressing the asymptotic behavior,
to fix the parameters of the approach data in the low-energy region say for center-of-mass energies
below $1-1.5\GeV$ will be most beneficial and are expected to have the largest potential impact
on the HLbL contribution.

\subsection{Pion transition form factor}

One of the central building blocks in Fig.~\ref{fig:flowchart} is $\pi\pi$
scattering, whose phase shifts, by virtue of Watson's final-state
theorem~\cite{Watson}, are required for the resummation of $\pi\pi$
rescattering corrections. The corresponding analyses of
$\omega,\phi\to3\pi$~\cite{NKS} and $\gamma\pi\to\pi\pi$~\cite{HKS} give
then access to the pion transition form factor with the isoscalar
virtuality either fixed to the mass of $\omega,\phi$ or to a real
isoscalar photon, respectively.  In particular, the formalism provides a
parametrization of $\gamma\pi\to\pi\pi$ that can be used to extract the
chiral $\gamma3\pi$ anomaly from data and thereby check the underlying
low-energy theorem. For general isoscalar virtualities the normalization of
the amplitude cannot be predicted within dispersion theory, but has to be
fitted to data for the $e^+e^-\to3\pi$ spectrum. Combining the isoscalar
and isovector channels allows for the confrontation with
$e^+e^-\to\pi^0\gamma$ data~\cite{TFF}.

In order to illustrate the predictive power of the dispersive
representation of the various amplitudes, we discuss the number of
subtractions in the program outlined above in some more detail.  Both
$\omega,\phi\to3\pi$ and $\gamma\pi\to\pi\pi$ are dominated by a single
partial wave (the $P$-wave), and standard arguments on a realistic
high-energy behavior suggest a single subtraction constant should in
principle be sufficient.  This is given by the chiral anomaly $F_{3\pi}$
for $\gamma\pi\to\pi\pi$ (and can be used as theoretical input in the
absence of a precise experimental extraction~\cite{HKS}), and can be
determined from the partial decay widths $\Gamma_{3\pi}$ of
$\omega,\phi\to3\pi$ for the decays~\cite{NKS}.  Such singly-subtracted
three-pion partial waves subsequently allow for an unsubtracted dispersion
relation for the corresponding transition form factors (with the
charged-pion vector form factor as its sole additional input); in
particular, sum rules exist for the decay widths $\Gamma_{\pi^0\gamma}$ of
$\omega,\phi\to\pi^0\gamma$~\cite{SKN} as well as for the chiral anomaly
$F_{\pi^0\gamma\gamma}$ for $\pi^0\to\gamma\gamma$~\cite{HKS}. A
representation of the corresponding unitarity relations, together with the
list of necessary and optional subtractions, is given in
Table~\ref{table:TFF}. The first panel refers to the process with vanishing
isoscalar virtuality $q_s^2=0$, the second to $q_s^2=M_\omega^2,M_\phi^2$,
and the third to the general case.

As all these dispersion relations are constrained to \emph{elastic}
unitarity, i.e.\ only take two-pion intermediate states (in the isovector
$P$-wave channel) into account, the accuracy of these is expected not to be
perfect, and indeed can be checked experimentally.  A high-statistics
Dalitz plot for $\phi\to3\pi$~\cite{KLOE:phi3pi} was shown to be described
perfectly only as soon as a second subtraction was introduced to improve
the convergence of the dispersive integrals, and to suppress inelastic
effects~\cite{NKS}.  Similarly, the theoretical amplitude to accurately
extract the $\gamma3\pi$ anomaly from data was also formulated as a
two-parameter, twice-subtracted representation for the cross section
$\sigma(\gamma\pi\to\pi\pi)$~\cite{HKS}.  The above-mentioned sum rules for
transition form factor normalizations are found to be saturated by two-pion
intermediate states at the 90\% level; very similar results were also found
for the (singly-virtual) $\eta$ transition form factor~\cite{Hanhart_eta}.
In the general case, a second subtraction could be implemented by
interpolating between $q_s^2=0$ and $q_s^2=M_\omega^2,M_\phi^2$ with a
representation analogous to the one used for the $e^+e^-\to3\pi$
spectrum~\cite{TFF}.

While improving on the accuracy of dispersive representations at low
energies, additional subtractions in general lead to less convergent
amplitudes in the high-energy limit.  In this sense, the number of
subtractions chosen for the $\gamma^*\to3\pi$ partial waves cannot be
considered independently of the dispersive representation for the
transition form factors constructed therewith: in principle, oversubtracted
partial waves for $e^+e^-\to3\pi$ also necessitate a further subtraction
for the transition form factors. For example, for a twice-subtracted
$\omega\to3\pi$ partial wave, there is no formally convergent sum rule for
the amplitude $\omega\to\pi^0\gamma$ (although the contribution to the
dispersive integral from the low-energy region below 1~GeV may in practice
differ very little).  In this sense, the two columns for indispensable
(SC~1) and optional (SC~2) subtractions in Table~\ref{table:TFF} are
required to be applied consistently (with the exception of the last line
concerning the parametrization of the $e^+e^-\to3\pi$ cross section, which
concerns a different kinematical variable, the three-pion invariant mass).

We expect to improve on the convergence of the dispersive calculation of
the $q^2$-dependence of the transition form factors when oversubtracting
them once. In this way, the singly-virtual $\pi^0$ transition form factor
as tested in $e^+e^-\to\pi^0\gamma$, in the last step, serves as input to
fix a subtraction function required for a reliable prediction of the full
doubly-virtual transition form factor. Complementary information extracted
from $e^+e^-\to e^+e^-\pi^0$ directly, either in the time-like or
space-like region, provides further checks of consistency and could
potentially improve the accuracy of the form-factor determination.

\begin{table*}[t!]
\centering
\renewcommand{\arraystretch}{1.3}
\begin{tabular}{cccc}
\toprule
process & unitarity relations & SC 1& SC 2\\
\midrule
\includegraphics[width=3cm]{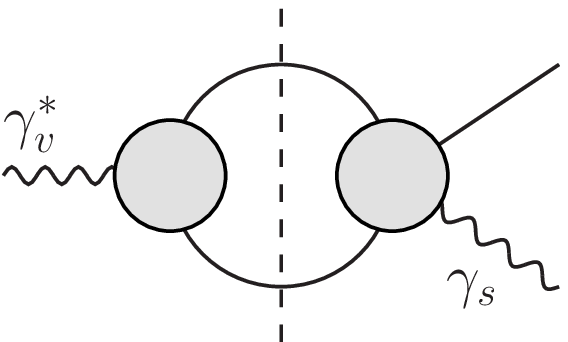} & \includegraphics[width=3cm]{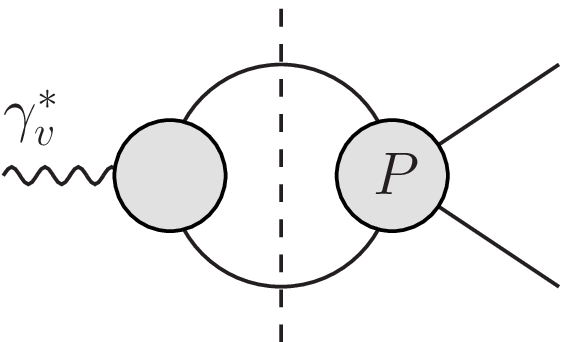} && \raisebox{0.8cm}{$F_{\pi^0\gamma\gamma}$}\\
& \includegraphics[width=3cm]{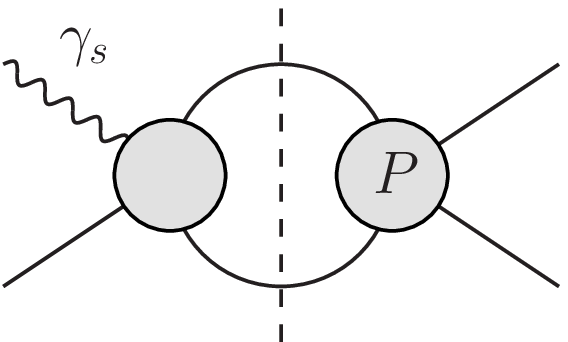} & \raisebox{0.9cm}{$F_{3\pi}$} & \raisebox{0.9cm}{$\sigma(\gamma\pi\to\pi\pi)$}\\
\midrule
\raisebox{0.15cm}{\includegraphics[width=3cm]{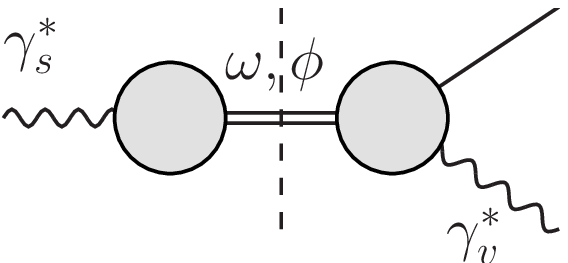}} & \includegraphics[width=3cm]{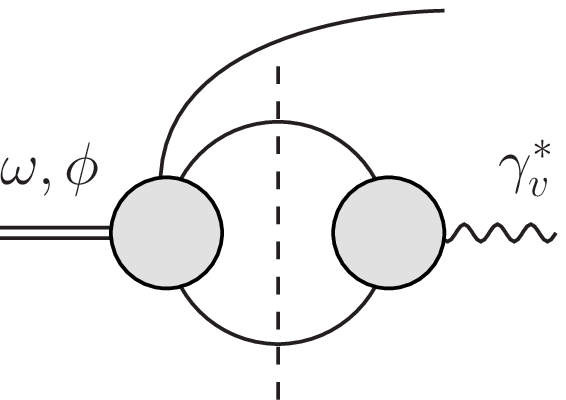} && \raisebox{0.9cm}{$\Gamma_{\pi^0\gamma}$}\\
& \includegraphics[width=3cm]{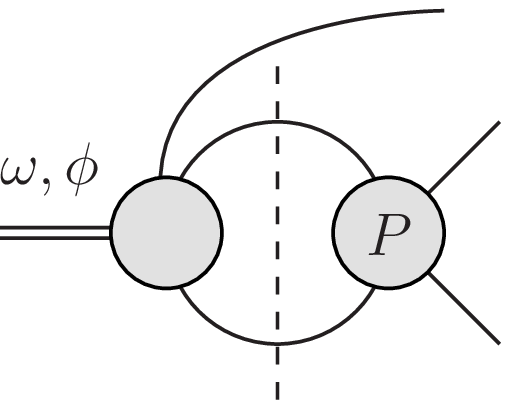} & \raisebox{0.9cm}{$\Gamma_{3\pi}$} & \raisebox{0.9cm}{$\frac{\diff^2 \Gamma}{\diff s\diff t}(\omega,\phi\to3\pi)$}\\
\midrule
\raisebox{0.3cm}{\includegraphics[width=2cm]{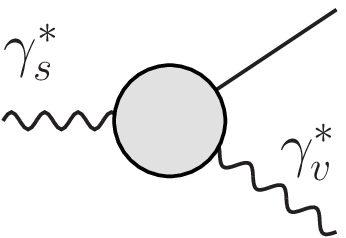}} & \includegraphics[width=3cm]{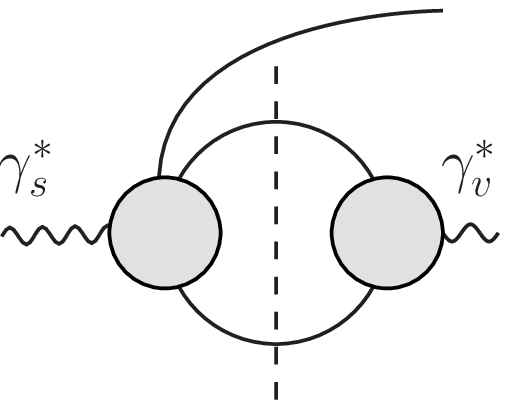} & & \raisebox{0.9cm}{$\sigma(e^+e^-\to \pi^0\gamma)$}\\
 & \includegraphics[width=3cm]{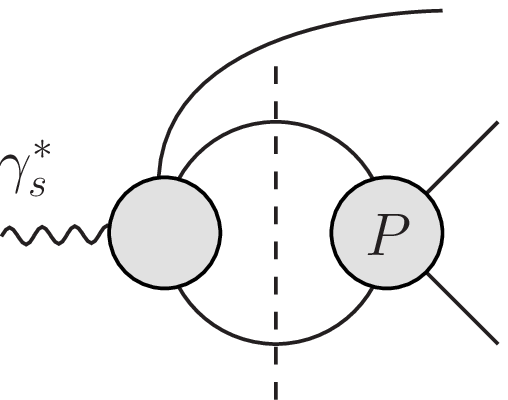} & \raisebox{1cm}{$\sigma(e^+e^-\to3\pi)$} & \raisebox{1cm}{$\begin{matrix} \sigma(\gamma\pi\to\pi\pi) \\ \frac{\diff^2 \Gamma}{\diff s\diff t}(\omega,\phi\to3\pi)\end{matrix}$} \\
  & \includegraphics[width=1.8cm]{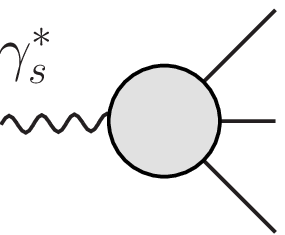} & \raisebox{0.7cm}{$F_{3\pi}$} & \raisebox{0.7cm}{$\sigma(e^+e^-\to3\pi)$}\\
\bottomrule
\end{tabular}
\renewcommand{\arraystretch}{1.0}
\caption{Processes and unitarity relations relevant for the pion transition form factor. 
The three panels represent $q_s^2=0$, $q_s^2=M_\omega^2,M_\phi^2$, and general $q_s^2$.
The last two columns refer to observables necessary to fix indispensable (SC~1) and  optional (SC~2) subtractions, respectively. $\gamma_{v/s}$ denotes isovector/isoscalar photons, capital letters the partial wave relevant for the $\pi\pi$ rescattering.
The last line is not formally a unitarity relation, but describes the parametrization of $\sigma(e^+e^-\to 3\pi)$.}
\label{table:TFF}
\end{table*}

\subsection{Partial waves for $\gamma^*\gamma^*\to\pi\pi$}

Several of the quantities mentioned in the context of the pion transition
form factor also feature prominently in the calculation of the helicity
partial waves for $\gamma^*\gamma^*\to\pi\pi$. First of all, $\pi\pi$
scattering determines their unitarity relation, i.e.\ their right-hand cut.
The leading contribution to the left-hand cut is generated by the pion
pole, with the coupling to the virtual photons described by the pion vector
form factor. Multi-pion contributions to the left-hand cut are usually
approximated in an effective resonance description, e.g.\ $2\pi$ would
correspond to the $\rho$ and $3\pi$ to $\omega,\phi$ (and, at higher
energies, to the axial-vector $a_1$). The widths of $\omega,\phi$ are
sufficiently small that a narrow-width approximation is justified, so that
the coupling to the virtual photons is governed by the $\omega,\phi$
transition form factors. In contrast, the width of the $\rho$ can be
strictly incorporated by expressing its contribution in terms of a
dispersive integral with spectral function determined by the $P$-wave for
$\gamma^*\pi\to\pi\pi$, one of the processes already appearing in the
context of the pion transition form factor. A representation of these
building blocks for the description of the left-hand cut is given in the
upper panel of Table~\ref{table:ggpipi}.\footnote{In this discussion, we
  confine ourselves to (multi-)pion intermediate states only. With the
  (isoscalar) photon virtuality at the $\phi$ mass, the left-hand cut will
  be dominated by kaon pole terms, rather than by two pions; see
  Sect.~\ref{sec:higher_IS} for the $\pi\pi/K\bar K$ coupled-channel
  system.}

An important aspect of the dispersive reconstruction of
$\gamma^*\gamma^*\to\pi\pi$ concerns subtraction
functions~\cite{GMM,HPS,Moussallam13}, at least one subtraction appears to
be necessary in most partial waves. In the on-shell case, the subtraction
constants may be identified with pion polarizabilities and either taken
from experiment or from Chiral Perturbation Theory (ChPT): one subtraction
requires knowledge of the dipole polarizabilities $\alpha_1\pm\beta_1$,
while a second subtraction involves also the quadrupole polarizabilities
$\alpha_2\pm\beta_2$. In general, however, the subtraction constants become
functions of the virtualities of the photons. As long as these virtualities
are small, the prediction from ChPT is available, but beyond the range of
validity of the chiral expansion data input is needed.
In~\cite{Moussallam13} data for $e^+e^-\to\pi\pi\gamma$ were used to fix
the subtraction function for the singly-virtual case. Again, the strength
of the dispersive approach is that this information collected in the
time-like region can be carried over to space-like kinematics, even though
for the doubly-virtual case control over anomalous thresholds is
crucial~\cite{HCPS}.  All data on $e^+e^-\to e^+e^-\pi\pi$, be it in the
real, singly-, or doubly-virtual case, would help constrain the subtraction
functions, for which, in turn, one may analyze a dispersive representation,
with subtraction constants fixed by ChPT and discontinuities by data on
$e^+e^-\to\pi\pi\gamma$ and $e^+e^-\to e^+e^-\pi\pi$. For the
singly-virtual case in the space-like region the subtraction function can
be expressed in terms of generalized pion polarizabilities
$\alpha_1\big(q^2\big)\pm\beta_1\big(q^2\big)$. Moreover, the
doubly-virtual subtraction functions are already partially constrained by
singly-virtual input, so that in combination with ChPT a first estimate
should be possible even absent doubly-virtual data.  These aspects are
represented by the lower panel of Table~\ref{table:ggpipi}.

\begin{table}[t!]
\centering
\renewcommand{\arraystretch}{1.3}
\begin{tabular}{cccc}
\toprule
process &  building blocks and SC\\
\midrule\\[-4mm]
\includegraphics[width=1.5cm,clip]{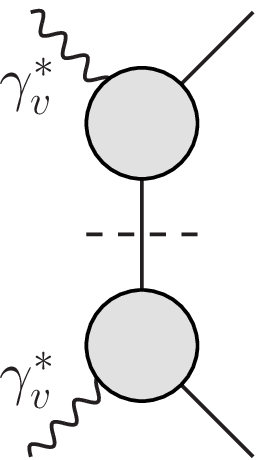} & \raisebox{0.7cm}{\includegraphics[width=1.5cm]{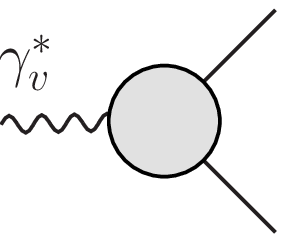}} \\[1mm]
\includegraphics[width=1.9cm]{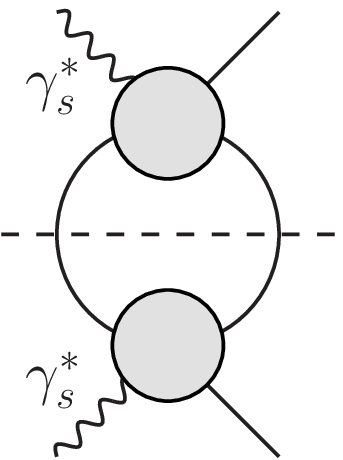} & \raisebox{0.7cm}{\includegraphics[width=1.5cm]{gstar3pi_vertex.eps}} \\[1mm]
\includegraphics[width=1.7cm]{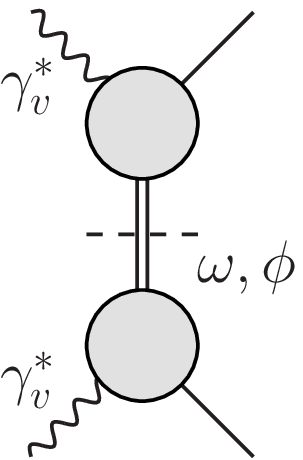} & \raisebox{0.7cm}{\includegraphics[width=1.7cm]{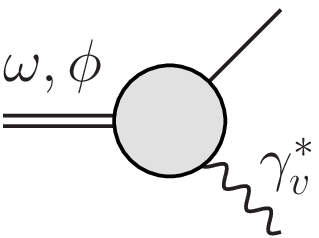}} \\
\midrule\\[-4mm]
\includegraphics[width=3.3cm,clip]{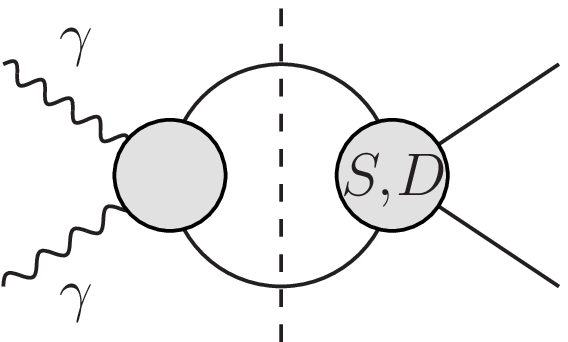}&\raisebox{0.8cm}{$\alpha_1\pm\beta_1$, $\alpha_2\pm\beta_2$}\\[1mm]
\includegraphics[width=3.3cm]{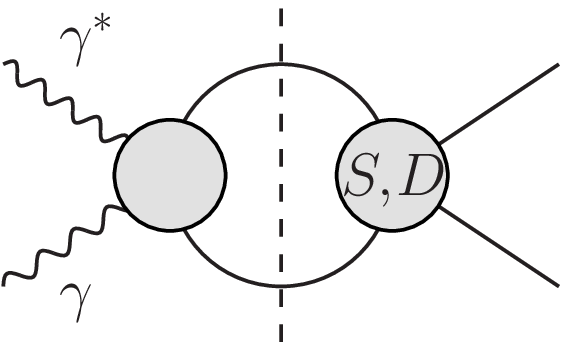}&\raisebox{0.9cm}{$\begin{matrix}\alpha_1\big(q^2\big)\pm\beta_1\big(q^2\big), \text{ChPT}\\
                                                 e^+e^-\to\pi\pi\gamma\\
                                                 e^+e^-\to e^+e^-\pi\pi
                                                \end{matrix}$}
\\[1mm]
\includegraphics[width=3.3cm]{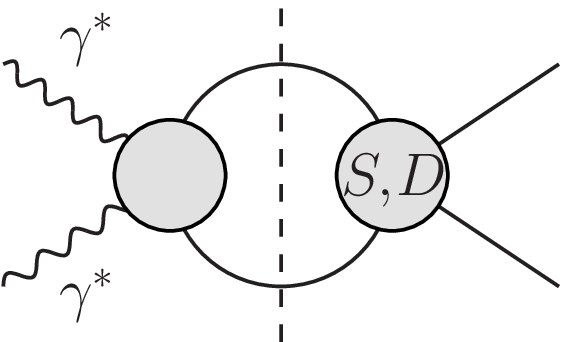}&\raisebox{0.9cm}{$\begin{matrix}\text{ChPT}\\
                                                 (e^+e^-\to\pi\pi\gamma)\\
                                                 e^+e^-\to e^+e^-\pi\pi
                                                \end{matrix}$}
\\
\bottomrule
\end{tabular}
\renewcommand{\arraystretch}{1.0}
\caption{Processes and unitarity relations relevant for the
  $\gamma^*\gamma^*\to\pi\pi$ partial waves. The last column refers to data
  that are required to determine sub-amplitudes and fix subtraction
  constants (SC). Processes in the upper panel are needed for the
  reconstruction of the left-hand cut, while the diagrams in the lower
  panel represent the right-hand cut in real, singly-virtual, and
  doubly-virtual $\gamma\gamma$ fusion. 
  In the doubly-virtual case, $e^+e^-\to\pi\pi\gamma$ is put in brackets
  since not all subtraction constants can be determined from singly-virtual
  data alone.}
\label{table:ggpipi}
\end{table}

\section{Higher intermediate states}
\label{sec:higher_IS}

Pseudoscalar poles with higher mass, most prominently $\eta$ and $\eta'$,
can be treated in the same way as sketched here for the pion pole, for
first steps in this direction see~\cite{Stollenwerk,Hanhart_eta}. As
alluded to in Sect.~\ref{sec:pion_pole}, in the end the result of the
dispersive calculation, valid in the low- and intermediate-energy region,
has to be brought into accord with constraints from perturbative QCD, which
can be interpreted as being generated by the exchange of even heavier
pseudoscalar resonances~\cite{MV}.

As far as two-particle intermediate states are concerned, the discussion
here generalizes immediately to $K \bar K$. In fact, in order to reproduce
the dynamics in the isospin-zero $S$-wave in the region of the $f_0(980)$
correctly, even a coupled-channel treatment of the $\pi\pi/K\bar K$ system
for this partial wave will become necessary. Further intermediate states,
e.g.\ with more than two pions, are more difficult to account for at the
same level of rigor as presented here. Possible approaches would be to
estimate effects of missing degrees of freedom in terms of an effective
resonance description~\cite{Pauk}, to cluster particles into effective
two-particle intermediate states for which a variant of~\eqref{amu_pole}
should exist, or to try to find a generalization of the FsQED calculation
including resonances, all of which concern possible future extensions of
the formalism.  The most important effective-resonance contributions not
coupling to $\pi\pi$ appear to be axial vectors $a_1(1260)$ and
$f_1(1285)$, as well as the scalar and tensor states of isospin $I=1$,
coupling to the $\pi\eta$/$K\bar K$ system, the $a_0(980)$ and $a_2(1320)$.
Sum-rule constraints relating different such resonance 
contributions~\cite{Pascalutsa:2012pr} should be taken into account where possible.
Apart from such estimates, clearly more work is needed to incorporate
constraints from perturbative QCD.

\section{Relation to previously considered contributions}
\label{sec:othermodels}

We wish to briefly comment on the relation of the dispersive analyses of
$\pi^0$-pole and $\pi\pi$-cut contributions to the HLbL tensor in the
context of previous analyses.

The pseudoscalar pole terms with their associated form factors are mostly
analyzed within vector-meson-dominance (VMD) models and extensions thereof~\cite{Bijnens:2001cq,Hayakawa:1997rq,KN,MV,Roig:2014uja,Masjuan:2012wy}.  
However, arguably the only experimental information we
have on the doubly-virtual $\pi^0$ transition form factor, via the
conversion decay $\omega\to\pi^0\mu^+\mu^-$, seems to indicate very
dramatic deviations from a simple VMD picture~\cite{NA60,NA60new}.  While
there are doubts about the consistency of these form factor data with what
is obtained, in a different kinematic regime, from
$e^+e^-\to\omega\pi^0$~\cite{SND-omegapi,CMD2-omegapi,KLOE:omegapi0gamma,SND-omegapi_2013},
enhancements in the (isovector) slope by more than 40\% are also found in a
theoretical dispersive description~\cite{SKN}.

The dispersive approach to include two-pion-cut contributions to the HLbL
tensor comprises various effects that have been discussed separately in the
literature.  It is the only sensible way to include the $f_0(500)$ scalar
meson, with its pole position deep inside the complex plane~\cite{CCL},
and---once generalized to a two-channel analysis including $K\bar
K$---certainly the cleanest for the $f_0(980)$, too.  Even the largest
tensor-meson effect, through the $f_2(1270)$~\cite{Pauk}, is covered by the
$\pi\pi$ $D$-wave contribution, as the $f_2(1270)$ is still dominantly
elastic.  Furthermore, the effects of pion polarizabilities on
HLbL~\cite{Engel1,Engel2} are automatically taken into account.

\section{Conclusions}

In this letter we have given an overview of recent theoretical developments
that will pave the way for a data-driven approach also to the 
calculation of the HLbL contribution to $a_\mu$. We have offered a detailed
account of which processes can help constrain the contribution from one-
and two-pion intermediate states to HLbL scattering. In particular, we
have discussed how information from other processes can provide a handle on the
dependence on the photon virtualities even in the absence of doubly-virtual
measurements, and specified the unitarity relations that are instrumental
in establishing this bridge. We are confident that with the methods
outlined here a more data-driven and thus less model-dependent evaluation
of the HLbL contribution to the muon $g-2$ is feasible.

\section*{Acknowledgements}
We would like to thank David W.\ Hertzog and Martin J.\ Savage for
discussions that prompted the preparation of this letter, as well as Simon
Eidelman and Andrzej Kup\'s\'c for comments on the manuscript.  Financial
support by the Swiss National Science Foundation, the DFG (CRC 16,
``Subnuclear Structure of Matter''), and by the project ``Study of Strongly
Interacting Matter'' (HadronPhysics3, Grant Agreement No.\ 283286) under
the 7th Framework Program of the EU is gratefully acknowledged.


\begin{thebibliography}{99}
\biboptions{sort&compress}

\bibitem{JN}
  F.~Jegerlehner and A.~Nyffeler,
  Phys.\ Rept.\ {\bf 477}  (2009) 1
  [arXiv:0902.3360 [hep-ph]].
  
\bibitem{Prades:2009tw}
  J.~Prades, E.~de Rafael and A.~Vainshtein,
  Advanced series on directions in high energy physics 20
  [arXiv:0901.0306 [hep-ph]].
  
\bibitem{g-2wp}
  T.~Blum {\it et al.},
  arXiv:1311.2198 [hep-ph].
  
\bibitem{Benayoun:2014tra}
  M.~Benayoun {\it et al.},
  arXiv:1407.4021 [hep-ph].
  
\bibitem{Calmet:1976kd}
  J.~Calmet, S.~Narison, M.~Perrottet and E.~de Rafael,
  Phys.\ Lett.\ B {\bf 61} (1976) 283.
  
\bibitem{Kurz:2014wya}
  A.~Kurz, T.~Liu, P.~Marquard and M.~Steinhauser,
  Phys.\ Lett.\ B {\bf 734} (2014) 144 
  [arXiv:1403.6400 [hep-ph]].

\bibitem{Colangelo:2014qya}
  G.~Colangelo, M.~Hoferichter, A.~Nyffeler, M.~Passera and P.~Stoffer,
  Phys.\ Lett.\ B {\bf 735} (2014) 90
  [arXiv:1403.7512 [hep-ph]].

\bibitem{CHPS}
  G.~Colangelo, M.~Hoferichter, M.~Procura and P.~Stoffer,
  arXiv:1402.7081 [hep-ph].
  
\bibitem{deRafael:1993za}
  E.~de Rafael,
  Phys.\ Lett.\ B {\bf 322} (1994) 239
  [hep-ph/9311316].

\bibitem{Bijnens:1995cc}
  J.~Bijnens, E.~Pallante and J.~Prades,
  Phys.\ Rev.\ Lett.\  {\bf 75} (1995) 1447
   [Erratum-ibid.\  {\bf 75} (1995) 3781]
  [hep-ph/9505251].
  
\bibitem{BPP95} 
  J.~Bijnens, E.~Pallante and J.~Prades,
  Nucl.\ Phys.\ B {\bf 474} (1996) 379
  [hep-ph/9511388].

\bibitem{Bijnens:2001cq}
  J.~Bijnens, E.~Pallante and J.~Prades,
  Nucl.\ Phys.\ B {\bf 626} (2002) 410
  [hep-ph/0112255].
 
\bibitem{Hayakawa:1995ps}
  M.~Hayakawa, T.~Kinoshita and A.~I.~Sanda,
  Phys.\ Rev.\ Lett.\  {\bf 75} (1995) 790
  [hep-ph/9503463].

\bibitem{Hayakawa:1996ki}
  M.~Hayakawa, T.~Kinoshita and A.~I.~Sanda,
  Phys.\ Rev.\ D {\bf 54} (1996) 3137
  [hep-ph/9601310].

\bibitem{Hayakawa:1997rq}
  M.~Hayakawa and T.~Kinoshita,
  Phys.\ Rev.\ D {\bf 57} (1998) 465
   [Erratum-ibid.\ D {\bf 66} (2002) 019902]
  [hep-ph/9708227 [hep-ph/0112102]].
 
\bibitem{Knecht:2001qg}
  M.~Knecht, A.~Nyffeler, M.~Perrottet and E.~de Rafael,
  Phys.\ Rev.\ Lett.\  {\bf 88} (2002) 071802
  [hep-ph/0111059].
  
\bibitem{KN}
  M.~Knecht and A.~Nyffeler,
  Phys.\ Rev.\ D {\bf 65} (2002) 073034
  [hep-ph/0111058].
  
\vfill\eject  
  
\bibitem{RamseyMusolf:2002cy}
  M.~J.~Ramsey-Musolf and M.~B.~Wise,
  Phys.\ Rev.\ Lett.\  {\bf 89} (2002) 041601
  [hep-ph/0201297].
  
\bibitem{MV}
  K.~Melnikov and A.~Vainshtein,
  Phys.\ Rev.\ D {\bf 70} (2004) 113006
  [hep-ph/0312226].

  \bibitem{Goecke:2010if}
  T.~Goecke, C.~S.~Fischer and R.~Williams,
  Phys.\ Rev.\ D {\bf 83} (2011) 094006
   [Erratum-ibid.\ D {\bf 86} (2012) 099901]
  [arXiv:1012.3886 [hep-ph]].
  
\bibitem{Roig:2014uja}
  P.~Roig, A.~Guevara and G.~L.~Castro,
  Phys.\ Rev.\ D {\bf 89} (2014) 073016
  [arXiv:1401.4099 [hep-ph]].
  
\bibitem{Blum:2014oka}
  T.~Blum, M.~Hayakawa and T.~Izubuchi,
  arXiv:1407.2923 [hep-lat].
  
\bibitem{Czerwinski}
  E.~Czerwi\'nski {\it et al.},
  arXiv:1207.6556 [hep-ph].
  
\bibitem{NKS}
  F.~Niecknig, B.~Kubis and S.~P.~Schneider,
  Eur.\ Phys.\ J.\ C {\bf 72} (2012) 2014
  [arXiv:1203.2501 [hep-ph]].
  
\bibitem{HKS}
  M.~Hoferichter, B.~Kubis and D.~Sakkas,
  Phys.\ Rev.\ D {\bf 86} (2012) 116009
  [arXiv:1210.6793 [hep-ph]].
  
\bibitem{SKN}
  S.~P.~Schneider, B.~Kubis and F.~Niecknig,
  Phys.\ Rev.\ D {\bf 86} (2012) 054013
  [arXiv:1206.3098 [hep-ph]].
 
\bibitem{MesonNet} 
  M.~J.~Amaryan {\it et al.},
  arXiv:1308.2575 [hep-ph].
 
\bibitem{GMM}
  R.~Garc\'ia-Mart\'in and B.~Moussallam,
  Eur.\ Phys.\ J.\ C {\bf 70} (2010) 155
  [arXiv:1006.5373 [hep-ph]].
 
\bibitem{HPS}
  M.~Hoferichter, D.~R.~Phillips and C.~Schat,
  Eur.\ Phys.\ J.\ C {\bf 71} (2011) 1743
  [arXiv:1106.4147 [hep-ph]].
  
\bibitem{Moussallam13} 
  B.~Moussallam,
   Eur.\ Phys.\ J.\ C {\bf 73} (2013)  2539
  [arXiv:1305.3143 [hep-ph]].
  
\bibitem{HCPS}
  M.~Hoferichter, G.~Colangelo, M.~Procura and P.~Stoffer,
  arXiv:1309.6877 [hep-ph].
  
\bibitem{Mandelstam} 
  S.~Mandelstam,
  { Phys.\ Rev.\ Lett.}\ {\bf 4} (1960) 84.
 
\bibitem{LMS} 
  W.~Lucha, D.~Melikhov and S.~Simula,
  Phys.\ Rev.\ D {\bf 75} (2007) 016001
  [hep-ph/0610330].

\bibitem{Watson}
  K.~M.~Watson,
  Phys.\ Rev.\  {\bf 95} (1954) 228.
  
\bibitem{TFF}
  M.~Hoferichter, B.~Kubis, S.~Leupold, F.~Niecknig and S.~P.~Schneider, in preparation.

\bibitem{KLOE:phi3pi}
  A.~Aloisio {\it et al.}  [KLOE Collaboration],
  Phys.\ Lett.\ B {\bf 561} (2003) 55
   [Erratum-ibid.\ B {\bf 609} (2005) 449]
  [hep-ex/0303016].

\bibitem{Hanhart_eta}
  C.~Hanhart, A.~Kup\'s\'c, U.-G.~Mei{\ss}ner, F.~Stollenwerk and A.~Wirzba,
  Eur.\ Phys.\ J.\ C {\bf 73} (2013) 2668
  [arXiv:1307.5654 [hep-ph]].
  
\bibitem{Stollenwerk}
  F.~Stollenwerk, C.~Hanhart, A.~Kup\'s\'c, U.-G.~Mei{\ss}ner and A.~Wirzba,
  Phys.\ Lett.\ B {\bf 707} (2012) 184
  [arXiv:1108.2419 [nucl-th]].
  
\bibitem{Pauk}
  V.~Pauk and M.~Vanderhaeghen,
  Eur.\ Phys.\ J.\ C {\bf 74} (2014) 3008
  [arXiv:1401.0832 [hep-ph]].
  
\bibitem{Pascalutsa:2012pr}
  V.~Pascalutsa, V.~Pauk and M.~Vanderhaeghen,
  Phys.\ Rev.\ D {\bf 85} (2012) 116001
  [arXiv:1204.0740 [hep-ph]].

\bibitem{Masjuan:2012wy}
  P.~Masjuan,
  Phys.\ Rev.\ D {\bf 86} (2012) 094021
  [arXiv:1206.2549 [hep-ph]].
  
\bibitem{NA60}
  R.~Arnaldi {\it et al.}  [NA60 Collaboration],
  Phys.\ Lett.\ B {\bf 677} (2009) 260
  [arXiv:0902.2547 [hep-ph]].

\bibitem{NA60new}
  G.~Usai  [NA60 Collaboration],
  Nucl.\ Phys.\  A {\bf 855} (2011) 189.

\bibitem{SND-omegapi}
  M.~N.~Achasov {\it et al.} [SND Collaboration],
  Phys.\ Lett.\ B {\bf 486} (2000) 29
  [hep-ex/0005032].

\bibitem{CMD2-omegapi}
  R.~R.~Akhmetshin {\it et al.}  [CMD-2 Collaboration],
  Phys.\ Lett.\ B {\bf 562} (2003) 173
  [hep-ex/0304009].

\bibitem{KLOE:omegapi0gamma}
  F.~Ambrosino {\it et al.}  [KLOE Collaboration],
  Phys.\ Lett.\ B {\bf 669} (2008) 223
  [arXiv:0807.4909 [hep-ex]].
  
\bibitem{SND-omegapi_2013}
  M.~N.~Achasov {\it et al.} [SND Collaboration],
  Phys.\ Rev.\ D {\bf 88} (2013) 5,  054013
  [arXiv:1303.5198 [hep-ex]].

\bibitem{CCL}
  I.~Caprini, G.~Colangelo and H.~Leutwyler,
  Phys.\ Rev.\ Lett.\  {\bf 96} (2006) 132001
  [hep-ph/0512364].

\bibitem{Engel1}
  K.~T.~Engel, H.~H.~Patel and M.~J.~Ramsey-Musolf,
  Phys.\ Rev.\ D {\bf 86} (2012) 037502
  [arXiv:1201.0809 [hep-ph]].

\bibitem{Engel2}
  K.~T.~Engel and M.~J.~Ramsey-Musolf,
  arXiv:1309.2225 [hep-ph].
  
\end{thebibliography}
\end{document}